\documentclass[10pt,twoside=false,captions=tableheading]{scrartcl}
\usepackage[utf8]{inputenc}

\usepackage{tempPaper}

\newcommand{\ie}{i.e.,\xspace}
\newcommand{\eg}{e.g.,\xspace}

\newcommand{\dFE}{\ensuremath{\bm{\hat{d}}}}
\newcommand{\uFE}{\ensuremath{\bm{\hat{u}}}}
\usepackage{here}

\DeclareRobustCommand\circled[1]{\tikz[baseline=(char.base)]{
            \node[shape=circle,draw=black,inner sep=0.8pt,fill=white] (char) {#1};}}

\DeclareAcronym{DOF}{
    short = DOF ,
    long = degree of freedom,
    short-plural = s,
    long-plural-form = degrees of freedom
}

\title{A hybrid multi-phase field model to describe cohesive failure in orthotropic materials, assessed by modeling failure mechanisms in wood}
\date{\today}
\author{Sebastian Pech, Markus Lukacevic, Josef Füssl}

\begin{document}
\maketitle

\section*{Abstract}

% One or two sentences providing a basic introduction to the field, comprehensible to a scientist in any discipline.
Fracture mechanics is crucial for many fields of engineering applications, as precisely predicting failure of structures and parts is required for efficient designs.
The simulation of failure processes is, both from a mechanical and a numerical point of view, challenging, especially for inhomogeneous materials, where the microstructure influences crack initiation and propagation and might lead to very complex crack patterns.
% Two to three sentences of more detailed background, comprehensible to scientists in related disciplines.
The phase field method for fracture is a promising approach to encounter such materials, since it is able to describe complex fracture phenomena like crack kinking, branching and coalescence.
Moreover, it is a largely mesh independent approach, given that the mesh is homogenous in the area of the crack.
However, the broadly used formulation of the phase field method is limited to isotropic materials and does not account for preferable fracture planes defined through the material's microstructure.
% One sentence clearly stating the general problem being addressed by this particular study
In this work, the method is expanded to take orthotropic constitutive behavior and preferable directions of crack propagation into account.
% One sentence summarising the main result (with the words “here we show” or their equivalent).
We show that by using a stress-based split and multiple phase field variables with preferable fracture planes, in combination with a hybrid phase field approach, a general framework can be found for simulating anisotropic, inhomogeneous materials.
% Two or three sentences explaining what the main result reveals in direct comparison to what was thought to be the case previously, or how the main result adds to previous knowledge.
The stress-based split is based on fictitious crack faces and is, herein, expanded to support anisotropic materials.
Furthermore, a novel hybrid approach is used, where the degradation of the sound material is performed based on a smooth traction free crack boundary condition, which proves to be the main driving factor for recovering commonly observed crack patterns.
% One or two sentences to put the results into a more general context.
This is shown by means of a detailed analysis of two examples: a wooden single edge notched plate and a wood board with a single knot and complex fiber directions.
In both cases, the proposed novel hybrid phase field approach is able to realistically reproduce complex failure modes.
% Two or three sentences to provide a broader perspective, readily comprehensible to a scientist in any discipline, may be included in the first paragraph if the editor considers that the accessibility of the paper is significantly enhanced by their inclusion. Under these circumstances, the length of the paragraph can be up to 300 words. (The above example is 190 words without the final section, and 250 words with it).

\section{Introduction}
\label{sec:Introduction}
The use of fracture mechanical modeling approaches is crucial in areas of applied engineering.
Being able to describe failure processes of structures and components realistically and allows for their optimization with regard to design and material consumption, while ensuring high reliability standards.
Simulation of failure processes has been the topic of numerous publications over the recent decades and is, due the complexity of possible material failure mechanisms, still an area that is being researched very intensively.
% - "Methods for simulating [[Fracture]] in solids #[[solid mechanics]] #[[📃 Zettel]]  #inprogress"

The foundation of most studies on brittle fracture processes is the work of \textcite{Griffith1921}, which describes fracture in terms of a critical fracture energy release rate required for crack propagation.
Based on Griffith's work, \textcite{Irwin1958} introduced the so-called stress intensity factors to characterize stress fields around the crack tip, depending on geometry and load.
While those linear elastic fracture mechanics theories can be applied to describe crack propagation of existing cracks, effects like crack kinking, branching, coalescence, initiation and cohesive behavior are not covered.
With the broad establishment of finite element methods for problems related to continuum mechanics, also new methods for simulating fracture processes emerged: 
Approaches based on remeshing and usage of special crack tip elements \parencite{barsoum1977,shahani2009}, the node split method \parencite{peng2012}, cohesive elements and cohesive zones models \parencite{Barenblatt1962,ponnusami2015} and XFEM \parencite{moes2002}.
Those methods allow overcoming some previously mentioned limitations of theories rooted in Griffiths's work, however, each approach comes with its own weaknesses.
% - "Origins of the [[Phase field]] method for fracture #Fracture #[[📃 Zettel]]"

One of the most recent and promising method is the so-called phase field method for fracture.
This method was initially proposed by \textcite{Francfort1998} and is also rooted in Griffith's theory of brittle fracture, however, formulated by a variational approach, through which the total energy of the system is minimized.
The main advantage is that no predefined crack paths are needed and branching as well as coalescence of cracks is naturally included in this approach.
However, finding a solution to the proposed framework turned out to be very difficult.
Therefore, in \textcite{Bourdin2000,Bourdin2008} a regularization method was developed that allows the minimization problem to be solved numerically efficient.
By introduction of an auxiliary field $d(x) \in [0,1]$ -- the so-called crack phase field -- the crack discontinuity is modeled by including a smooth transition zone from intact ($d = 0$) to cracked ($d = 1$) solid.
The width of this transition zone is controlled by a regularization or length scale parameter.
As this parameter approaches zero (\ie recovering the discontinuous transition from solid to crack), the solution gamma-converges to Griffith's theory.

% - "Mixed Mode failure in wood #[[Mixed Mode]] #failure #Wood #[[📃 Zettel]]"

By expressing cracks in form of a field variable, the mentioned complex fracture phenomena like kinking, branching and coalescence, naturally arise from the defining system of differential equations.
Thus, the phase field method theoretically allows crack topologies of arbitrary complexity, only limited by the mesh size and mesh structure.
This motivates usage of the phase field method for materials with a complex micro- or macrostructure, like concrete, fiber-reinforced composites, polycrystalline structures and wood.
In those materials, the micro- and macrostructure strongly affects both the elastic behavior, in the sense of having anisotropic constitutive relationships, and the crack topology, by introduction of favorable fracture planes due to \enquote{weak} principal material directions.

\subsection{Fracture phenomena of wood}

For wood, this results in crack topologies driven by both the direction of least resistance, orthogonal to the wood fiber direction, and the maximum principal stress \parencite{Smith2003}.
This causes the often observed zig-zag pattern, where cracks jump from one growth layer to another, representing a combination of Mode-I, Mode-II and Mode-III failure modes.
% - "[[Fracture]] at critical load level in [[Wood]] #[[📃 Zettel]] #[[critical load]] #[[stress at failure]]"
In this way, materialspecific microstructural features might influence the fracture behavior and crack propagation during and after crack formation in complex materials.
In wood, crack growth is mainly triggered by defects in the cell wall material at a microscopic level.
This induces the often observed decrease in macroscopic stiffness, evidently visible in load-deflection plots, \ie the nonlinear behavior before reaching the peak load. 
Close to the actual peak load the microscopic cracks localize and the actual macroscopic crack and fracture process zone forms.
Failure processes in wood after crack initiation, like cohesive behavior, where studied by \textcite{vasic2002} and are also observable in the experimental study of \textcite{Dourado2008}.
Their research concluded with identifying so-called fiber bridging as the main cause of toughening effects at the crack tip.
% This leads to the discrepancy of the fracture energy release rate and the specific fracture energy.

\subsection{The phase field method for anisotropic materials}

The phase field models proposed by \textcite{Amor2009, Miehe2010b} contain the assumptions of isotropic constitutive behavior and ideal-brittle fracture.
However, in recent years, three approaches to allow for consideration of anisotropic behavior were published:
\textcite{BLEYER2018213} proposed a method that introduces additional phase field \acp{DOF}, which are uncoupled in the geometrical terms of the phase field equation and thus allow for different fracture energy release rates and length scale parameters.
The actual coupling is introduced on a constitutive level.
\textcite{TEICHTMEISTER20171} presented a different approach for anisotropic fracture by adding preferable directions through a second order tensor -- the so-called structural tensor.
This tensor scales the gradient of the crack phase field and imposes an orientation on the geometrical terms of the phase field equation.
A similar method was pursued by \textcite{NGUYEN2017279}.
Here, a model with additional phase field \acp{DOF} is used where each of the phase variables is linked to its own second order structural tensor that invalidates crack growth in certain direction.
Effectively, by suppressing one direction, favorable fracture planes, defined by the microstructure of the material, can be considered.
The tensor used in their work is similar to the transversely isotropic formulation found in the work of \textcite{TEICHTMEISTER20171}.

\subsection{Cohesive behavior in the phase field method}

Already the original publication \parencite{Bourdin2000}, covering the numerical implementation of the model from \textcite{Francfort1998}, discusses the role of the length scale parameter.
This was further elaborated by \textcite{Amor2009}.
Originally, the length scale parameter was seen as a purely numerical value in the context of the regularization scheme.
However, due to the similarities of the phase field method with gradient damage approaches, considering it a material parameter is not far-fetched.
As already outlined, with this parameter tending towards zero, the solution gamma-converges to Griffith's theory, describing ideal brittle failure.
By increasing this parameter, more complex softening behavior can be described, and ductile effects of quasi-brittle material failure can be implemented
An alternative approach is proposed by \textcite{Wu2018a} in form of the so-called non-standard phase field method, where the phase field's characteristic functions are tuned to a cohesive zone model with a predefined traction separation constitutive relation.
In their work, the length scale parameter is still regarded as a purely numerical value.
However, including the material's tensile strength into the formulation allows controlling the ductile behavior.

\subsection{Scope of this work}

In this paper, a cohesive phase field implementation is presented based on the above-mentioned non-standard model from \textcite{Wu2018a}.
Furthermore, the outlined methods for considering anisotropic behavior are evaluated and their possible combinations and limitations are discussed.
Based on the work of \textcite{hu2020PhasefieldModelFracture}, a novel hybrid approach \parencite{Ambati2014}, employing a crack boundary condition for the degradation part of the phase field equation, is formulated.

The paper is organized as follows:
\Cref{sec::Fundamentals_and_methods} provides the theoretical background and introduces methods and concepts from recent literature.
Additionally, a generalization of the stress split from \textcite{steinke2018} for anisotropic materials is given and the hybrid approach, utilizing a crack boundary condition, is described.
In \Cref{sec:results_and_discussion} the introduced phase field models are evaluated using two numerical examples with different levels of complexity.
First, a simple single edge notch plate with a predefined material direction is looked at.
The simulation results serve as a basis for the discussion of the hybrid approach described in this work.
Secondly, a more complex realistic model of a wooden board with a single knot is simulated to show the effects of spatially varying principal material directions, resulting from different wood fiber orientations.
The paper closes with a short summary of the introduced concepts for applying the phase field model to orthotropic materials showing different complex damage mechanisms, and gives a brief outlook and concluding remarks.

\section{Fundamentals and Methods}
\label{sec::Fundamentals_and_methods}
\subsection{Fundamentals\label{sec:fundamentals}}

\begin{figure}[htb]
    \centering
    \includegraphics{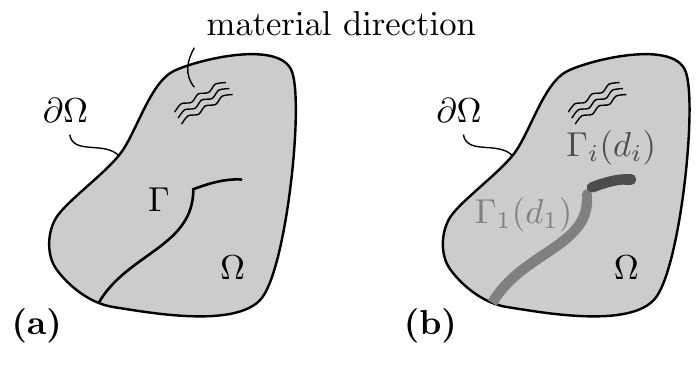}
    \caption{
        \textbf{(a)} sharp and \textbf{(b)} diffuse representation of a crack topology.
        Within the diffuse representation, the sharp crack $\Gamma$ in the body $\Omega$ is approximated using multiple diffuse crack fields, which depend on the crack phase fields $\bm{d}$.
    }
    \label{fig:Phasefield-regularization}
\end{figure}

To verify the various methods able to take anisotropic and quasi-brittle behavior into account, the unified phase field model from \textcite{Wu2018} is generalized to include $n$ different phase field variables \parencite{BLEYER2018213} and a second order structural tensor $\bm{A}_i$ \parencite{TEICHTMEISTER20171,NGUYEN2017279}, where $i$ denotes the $i$-th phase field.
Excluding body forces and surface tractions, the regularized form of the total energy $\Pi$ of the system, defined on the domain $\Omega$, thus reads:
\begin{align}
    \Pi(\bm{u},\bm{d}) = \int_\Omega\left[ \psi^+(\bm{u},\bm{d}) + \psi^-(\bm{u}) \right]d\Omega + \sum_i^n G_{c,i} \int_\Omega \gamma_i(d_i)d\Omega,
    \label{eq:phasefieldeq}
\end{align}
where $\bm{u}$ is the displacement field, $\bm{d}$ the phase field of dimensionality $n$, $G_{c,i}$ the critical energy release read of phase field $i$ and $\gamma_i$ is the regularized crack surface density functional that approximates the sharp crack surface.
As the driving force for crack propagation is energy based, a criterion is needed for preventing fracture under pure compressive stress states modes.
Thus, the strain energy density is separated into $\psi^+(\bm{u},\bm{d})$ which contributes to fracture and $\psi^-(\bm{u})$ which does not.
Only $\psi^+$ depends on the crack phase field $\bm{d}$, in such a way that the so-called degradation function $\omega(d_i)$, which damages the solid $\Omega$, degrades the strain energy density.

Within the unified phase field theory, the crack surface density functional is defined as
\begin{align}
    \gamma_i(d_i) = \frac{1}{c_{0,i}}\left[ \frac{1}{l_i} \alpha(d_i) + l_i\nabla{}d_i\bm{A}_i\nabla{}d_i \right]\;\text{with}\; c_{0,i} = 4\int_0^1\sqrt{\alpha(\zeta)}d\zeta.
\end{align}
As outlined in \textcite{Kuhn2015, Pham2010}, the function $\alpha(d_i)$ defines the local part of the dissipated fracture energy and determines the ultimate crack phase field.
It satisfies the properties $\alpha(d=0) = 0$ and $\alpha(d=1) = 1$ \parencite{Wu2018}.
In order to recover the actual surface measure of the crack set for $l_i \to 0$, the normalization constant $c_{0,i}$ is needed.
$\bm{A}_i$ is the so-called structural tensor, which scales the gradient of the crack phase field to define preferable or invalid crack propagation directions.
In a multi-phase field setting, the definition
\begin{align}
    \bm{A}_i = \bm{I} + \beta_i\left(\bm{I} - \bm{a}_i\otimes\bm{a}_i \right)\label{eq:structural_tensor}
\end{align}
from \textcite{NGUYEN2017279} can be used, which allows assigning material directions $\bm{a}_i$ and penalty factors $\beta_i$ for penalizing planes not orthogonal to the material directions for each phase field variable $d_i$.
For $\beta_i = 0$, the standard isotropic formulation of the crack surface density functional is recovered.

The two functions that mainly influence the fracture process are the degradation function $\omega(d_i)$ and $\alpha(d_i)$.
As described by \textcite{BLEYER2018213}, while multiple phase field variables are geometrically uncoupled (\eg $\alpha(d_i)$ is defined for each \ac{DOF}), they are coupled in the constitutive relation (\eg $\psi^+(\bm{u},\bm{d})$ is defined for all \acp{DOF}).
% Options for defining this tie will be discussed in \Cref{sec:facture_contributing_and_passive_parts}.
A general expression for those two functions is given by \textcite{Wu2017}, in the following way
\begin{align}
    \alpha_i(d_i) &= \xi d_i +(1-\xi) d_i^2 \quad \forall d_i \in [0,1]\;\xi \in [0,2] \;\text{and} \label{eq:localpart}\\
    \omega_i(d_i) &= \frac{(1-d_i)^p}{(1-d_i)^p+Q(d_i)} \quad p \geq 2 \label{eq:degrad} \\
    Q_i(d_i) &= a_{1,i}d_i + a_{1,i}a_{2,i}d_i^2 + a_{1,i}a_{2,i}a_{3,i}d_i^3, \label{eq:Q_a1_a2_a3}
\end{align}
where $a_{1,i}$, $a_{2,i}$ and $a_{3,i}$ are coefficients that can be calibrated to model a certain cohesive behavior related to the $i$-th phase field.

The general expression for the local part of the dissipated fracture energy resembles the well-known monotonous model.
Alternative formulations are of type of a double well function.
Such a formulation has an interesting property, because it provides an energy barrier between the undamaged and the fully damaged state, and thus, naturally models irreversibility of the crack phase field \parencite{Kuhn2015}.
However, those functions are barely used because of the two states being energetically equivalent, which can lead to phase fields expanding perpendicular to the crack path.
\Cref{eq:localpart} can be specialized for the two commonly used models $\alpha(d_i) = d_i$ for $\xi = 1$ (AT-1) and $\alpha(d_i) = d_i^2$ for $\xi = 0$ (AT-2) \parencite{ambrosio1990}.

As outlined in \textcite{Miehe2010b}, the degradation function must satisfy $\omega(d_i = 0)=1$ and $\omega(d_i = 1)=0$.
Furthermore, as the first derivative of the degradation function with respect to the phase field variable controls the amount of energetic driving force, $\omega'(d_i = 1)=0$ is needed, in order to eliminate this elastic driving term once full damage is reached \parencite{Miehe2010b,Kuhn2015}.
This ultimately stops further crack growth in regions characterized by $d_i = 1$.
\textcite{steinke2018} discuss the additional soft requirement of $\omega(d_i = 0) \neq 0$, which, if not satisfied, as well leads to the elimination of the elastic driving term.
For $d_i = 0$, \ie the undamaged state, crack growth is hindered, and no phase field evolution can take place.
Therefore, such a model requires additional treatment in form of a numerical perturbation of the initial state, such that the energetic driving forces become unequal to zero.
One commonly used definition is $\omega(d_i) = (1-d_i)^2$ \parencite{Miehe2010}.
The general expression in \Cref{eq:degrad} contains this simple case for $p=2$, $a_1=2$, $a_2=-\sfrac{1}{2}$ and $a_3=0$.

The remaining part left to be specified from \Cref{eq:phasefieldeq} is the strain energy density split.
In order to properly discuss the various kinds of methods to approach this separation, at first, two commonly applied methods for dealing with coupled equations within the variation framework are discussed.

\subsection{Isotropic, anisotropic and hybrid formulation\label{sec:isotropic_anisto}}

As described by \textcite{Ambati2014}, there are two basic formulations originating from the regularized variational framework \parencite{Bourdin2000,Bourdin2008}, the isotropic formulation and the anisotropic formulation \parencite{Miehe2010b, Miehe2010}\footnote{The terms isotropic and anisotropic are not to be interpreted in terms of the local material behavior, but they refer to the decomposition of the strain energy density}.
The isotropic formulation does not contain the additive decomposition of the strain energy density and, thus, gives a linear relation in $\bm{u}$, reading
\begin{align}
    \bm{\sigma}(\bm{u},\bm{d}) = \frac{\partial\psi(\bm{u},\bm{d})}{\partial\bm{\varepsilon}}.
    \label{eq:constiso}
\end{align}
The anisotropic formulation
\begin{align}
    \bm{\sigma}(\bm{u},\bm{d}) = \frac{\partial \psi^+(\bm{u},\bm{d}) + \psi^-(\bm{u}) }{\partial\bm{\varepsilon}},
    \label{eq:constaniso}
\end{align}
however, contains the split and is thus non-linear in $\bm{u}$.
This property simplifies the solution process when the so-called staggered approach is used (see \Cref{sec:Solver}), as the deformation subproblem can be treated as an uncoupled linear problem.

One obvious downside of the isotropic formulation is that every deformation state is degraded equally, thus unphysical behavior like interpenetration of crack faces or crack growth under pure compressive stress states can occur.
Each effect is related to a different aspect of the evolution equations.
Crack growth is rooted in the energetic driving force and interpenetration of crack faces in the constitutive relation.
The so-called hybrid formulation \parencite{Ambati2014} combines the advantages of both formulations, \ie the linear behavior of the isotropic formulation and the physical, more appropriate, modeling of the anisotropic one.
This is achieved by using the constitutive relation from \Cref{eq:constiso} with the additional constraint that for all $\bm{u}$ with $\psi^+$ smaller than $\psi^-$ (\ie the passive energy parts outweighs the crack driving one), the material is treated as undamaged, \ie $d_i = 0$.
For the energetic driving force, the anisotropic formulation is used, leading to the proper crack propagation behavior.
As shown in \textcite{Ambati2014}, the hybrid formulation manages to produce results qualitatively and quantitatively similar to the anisotropic formulation, with a computation effort close to the isotropic formulation.

\subsection{Fracture contributing and passive parts\label{sec:facture_contributing_and_passive_parts}}

The energy split into a part that drives and is affected by fracture and into a part that is neither affected by nor drives fracture is one key ingredient of the phase field formulation for ensuring a realistic fracture behavior.
As outlined in \Cref{sec:isotropic_anisto}, by ignoring this split, \ie using the isotropic formulation, effects like interpenetration of crack faces or crack growth under pure compressive stress states can occur.
This unphysical behavior can be eliminated by properly splitting the energy, such that the energetic driving force and also the damaged part in the constitutive relation is only related to the contributing part $\psi^+$.

Besides this physical motivation, there is also a simple conceptional one: The additive decomposition $\psi = \psi^+ + \psi^-$, must retain its validity.
In \textcite{dijk2020}, this is very well explained based on the two common methods for splitting the strain energy density: the spectral decomposition by \textcite{Miehe2010b} and the volumetric-deviatoric decomposition by \textcite{Amor2009}.
Starting with the well-known formulation of the strain energy density, the fundamental idea is to separate stresses and strains into strictly fracture-contributing and passive parts.
Hence, the strain energy density reads
\begin{align}
    \psi &= \frac{1}{2} (\bm{\sigma}^+ + \bm{\sigma}^-) : (\bm{\varepsilon}^+ + \bm{\varepsilon}^-)\label{eq:standard_decompo}\\
         &= \frac{1}{2} ( \underbrace{\bm{\sigma}^+ : \bm{\varepsilon}^+}_{\psi^+} + \bm{\sigma}^+ : \bm{\varepsilon}^- + \bm{\sigma}^- : \bm{\varepsilon}^+ + \underbrace{\bm{\sigma}^- : \bm{\varepsilon}^-}_{\psi^-}).
\end{align}
So, for the additive decomposition to hold, the terms $\bm{\sigma}^+ : \bm{\varepsilon}^-$ and $\bm{\sigma}^- : \bm{\varepsilon}^+$, consisting of contributing and passive parts, must vanish.

The spectral decomposition is conceptually rooted in fracture mechanics, so tensile principal stresses (and strains) result in crack opening, while compressive principal stresses (and strains) do not.
This idea motivates the decomposition to be performed in the principal strain space.
There, for an isotropic material, where the directions of the principal strains and principal stresses coincide, also the contributing strains are orthogonal to the passive stresses and the passive strains are orthogonal to the contributing stresses.
Therefore, as required, the terms containing both contributing and passive parts indeed vanish.
By the same reasoning, the volumetric-deviatoric decomposition is also valid for an isotropic material, as the volumetric and deviatoric parts are orthogonal.
So both models can be applied to isotropic materials.
However, for non-isotropic materials, this orthogonality between principle stresses and strains no longer exists and, thus, neither of the above models can be used, as $\psi \neq \psi^+ + \psi^-$.

The phase field formulation consists of two essential parts, one being the formulation of the energetic driving force and the other being the actual constitutive behavior.
On those two parts, the above described decomposition has a very different impact.
Regarding the first part, \textcite{Miehe2015} introduced the concept of a dimensionless crack driving function, which is a generalization of the history function \parencite{Miehe2010b} and allows replacing the energetic driving force by an arbitrary failure function, like a maximum stress, a maximum strain or a Tsai-Wu \parencite{GULTEKIN201823} criterion.
While this opens the phase field formulation to a variety of materials, the restrictions on the constitutive relation remain.
However, the current literature finally provides three approaches of a valid strain energy decomposition for non-isotropic materials:
\begin{itemize}
    \item A proper split for anisotropic materials \parencite{dijk2020}, based on generalizations of decomposition from \textcite{Miehe2010b} and \textcite{Amor2009},  which is not yet validated,
    \item a stress-based split \parencite{steinke2018, hu2020PhasefieldModelFracture} that is based on a different formulation of \Cref{eq:standard_decompo} and
    \item the hybrid approach described in \Cref{sec:isotropic_anisto}.
\end{itemize}

The hybrid approach is a simple method to circumvent the validity of the split, as for the constitutive part no split is required.
In a sense, the hybrid approach represents a generalized framework for applying arbitrary crack driving functions, as long as an additional constraint ensures that crack faces cannot interpenetrate.
The additional constraint formulation from \textcite{Ambati2014}, where for $\psi^- > \psi^+$ contact of the crack faces is assumed, however, might not be suitable for any crack driving function, thus, requiring an alternative formulation of this constraint, suited for ensuring a physical, sound fracture behavior.

A stress-based split represents an alternative approach for separating the strain energy density into a contributing and a passive part.
The method, as described by \textcite{steinke2018}, is based on the idea that only stresses are additively decomposed.
So \Cref{eq:standard_decompo} changes to
\begin{align}
    \psi &= \frac{1}{2} (\bm{\sigma}^+ + \bm{\sigma}^-) : (\bm{\varepsilon}) = \frac{1}{2} ( \underbrace{\bm{\sigma}^+ : \bm{\varepsilon}}_{\psi^+} + \underbrace{\bm{\sigma}^- : \bm{\varepsilon}}_{\psi^-}).
\end{align}
Obviously, this approach is applicable to any constitutive behavior, as there are no terms consisting of both contributing and passive parts.
Nevertheless, \textcite{steinke2018} specialized the model for isotropic materials, as some observations on the inherent properties of the deformations in presence of a crack, discussed below,  require changes in the initial formulation of $\bm{\sigma}^+$.

\begin{figure}[htb]
    \centering
    \includegraphics{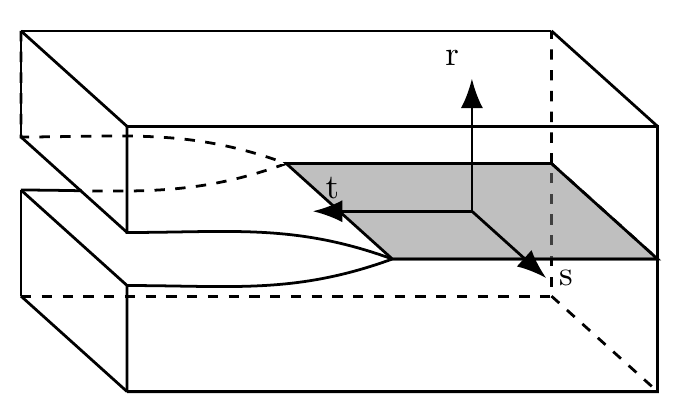}
    \caption{Crack coordinate system of a fictitious crack surface, required to identify Mode-I, Mode-II and Mode-III crack driving forces.}
    \label{fig:CSS}
\end{figure}

The stress based split is performed in a local crack coordinate system, as shown in \Cref{fig:CSS}.
This coordinate system related to a fictitious crack surface allows identification of crack driving forces for Mode-I, Mode-II and Mode-III fracture.
By representing the stress tensor in the crack coordinate system, the crack driving and passive stresses can be identified:
\begin{align}
    \bm{\sigma}^+ &= \underbrace{\left<\sigma_{rr}\right>_{+}(\bm{r}\otimes\bm{r})}_{\text{Mode-I}} + \underbrace{\bm{\sigma}_{rt} + \bm{\sigma}_{tr}}_{\text{Mode-II}} + \underbrace{\bm{\sigma}_{rs} + \bm{\sigma}_{sr}}_{\text{Mode-III}}\label{eq:stress_modes_1}\\
    \bm{\sigma}^- &= \left<\sigma_{rr}\right>_{-}(\bm{r}\otimes\bm{r}) + \bm{\sigma}_{tt} + \bm{\sigma}_{ss} + \bm{\sigma}_{ts} + \bm{\sigma}_{st},\label{eq:stress_modes_2}
\end{align}
where $\bm{\sigma}_{ij} = (\bm{\sigma}:(\bm{i}\otimes\bm{j}))(\bm{i}\otimes\bm{j})$, \ie the contributions to the stress tensor related to the $i$ and $j$ direction in the crack coordinate system and $\left<\bullet\right>_+ = \sfrac{(\bullet + |\bullet|)}{2}$ and $\left<\bullet\right>_- = \sfrac{(\bullet - |\bullet|)}{2}$.
At this point, the decomposition is still applicable to any constitutive law, however, it can lead to physically inconsistent results, as in the fully damaged state the essential crack boundary conditions -- no positive normal stress perpendicular to the crack and no shear stresses along a frictionless crack surface \parencite{strobl2015NovelTreatmentCrack} -- are not recovered for certain strain states.

This unphysical behavior can easily be shown by picturing a state of pure crack normal strain (\ie $\varepsilon_{rr} > 0$ and all other strain components equal to zero).
As a generalization of the isotropic model from \textcite{steinke2018}, following a linear elastic constitutive law, such a strain state would result in non-zero stress components:
\begin{align}
    \sigma_{rr} &= C_{rrrr}\varepsilon_{rr}\text{ and }\label{eq:normal_strain_1}\\
    \sigma_{ij} &= C_{ijrr}\varepsilon_{rr}.\label{eq:normal_strain_2}
\end{align}
A fully developed crack state, however, should be stress-free due to this very strain state, as the two crack surfaces should be able to move freely along the crack's normal vector.
Additionally, the stresses related to \emph{Poisson's} effect (\Cref{eq:normal_strain_2}), must be considered in the fracture contributing stresses $\bm{\sigma}^+$, by expressing them in terms of the crack normal stresses from \Cref{eq:normal_strain_1,eq:normal_strain_2} as
\begin{equation}
    \sigma_{ij} = \frac{C_{ijrr}}{C_{rrrr}}\sigma_{rr},
\end{equation}
which leads to the following formulation of the crack contributing and passive stresses:
\begin{align}
    \begin{split}
    \bm{\sigma}^+ &= \left<\sigma_{rr}\right>_{+}(\bm{r}\otimes\bm{r}) + \bm{\sigma}_{rt} + \bm{\sigma}_{tr} + \bm{\sigma}_{rs} + \bm{\sigma}_{sr} + \frac{\left<\sigma_{rr}\right>_{+}}{C_{rrrr}} \big[
        C_{ssrr}\cdot(\bm{s}\otimes\bm{s})+\\
        &+C_{ttrr}\cdot(\bm{t}\otimes\bm{t})+
        C_{rtrr}\cdot(\bm{r}\otimes\bm{t})+
        C_{rsrr}\cdot(\bm{r}\otimes\bm{s})+
        C_{trrr}\cdot(\bm{t}\otimes\bm{r})+
        C_{srrr}\cdot(\bm{s}\otimes\bm{r})+\\
        &+C_{tsrr}\cdot(\bm{t}\otimes\bm{s})+
        C_{strr}\cdot(\bm{s}\otimes\bm{t})\big]
    \end{split}\label{eq:total_active_stress}\\
    \begin{split}
    \bm{\sigma}^- &= \left<\sigma_{rr}\right>_{-}(\bm{r}\otimes\bm{r}) + \bm{\sigma}_{tt} + \bm{\sigma}_{ss} + \bm{\sigma}_{ts} + \bm{\sigma}_{st} - \frac{\left<\sigma_{rr}\right>_{+}}{C_{rrrr}} \big[
        C_{ssrr}\cdot(\bm{s}\otimes\bm{s})+\\
        &+C_{ttrr}\cdot(\bm{t}\otimes\bm{t})+
        C_{rtrr}\cdot(\bm{r}\otimes\bm{t})+
        C_{rsrr}\cdot(\bm{r}\otimes\bm{s})+
        C_{trrr}\cdot(\bm{t}\otimes\bm{r})+
        C_{srrr}\cdot(\bm{s}\otimes\bm{r})+\\
        &+C_{tsrr}\cdot(\bm{t}\otimes\bm{s})+
        C_{strr}\cdot(\bm{s}\otimes\bm{t})\big]
    \end{split}\label{eq:total_passive_stress}
\end{align}

\Cref{eq:total_active_stress} is based on \Cref{eq:stress_modes_1}, which contains crack driving stresses identified using classic fracture mechanics failure modes.
In addition, \Cref{eq:total_active_stress} also considers the stresses related to \emph{Poisson's} effect, which must vanish for a fully developed phase field.
Having this formulation for $\bm{\sigma}^+$ allows finding $\bm{\sigma}^-$ from $\bm{\sigma}=\bm{\sigma}^+ + \bm{\sigma}^-$.
It can be shown that this generalized formulation of the stress split contains the isotropic formulation proposed by \textcite{steinke2018}.
In the isotropic case, only the entries $C_{rrrr} = \lambda + 2 \mu$, $C_{ssrr} = C_{ttrr} = \lambda$, required for the additional decomposed stresses, are non-zero, leading to the expression
\begin{equation}
    \bm{\sigma}^+ = \left<\sigma_{rr}\right>_{+}(\bm{r}\otimes\bm{r}) + \bm{\sigma}_{rt} + \bm{\sigma}_{rs} + \bm{\sigma}_{tr} + \bm{\sigma}_{ts} + \frac{\lambda}{\lambda + 2 \mu}\left<\sigma_{rr}\right>_{+}(\bm{s}\otimes\bm{s} + \bm{t}\otimes\bm{t}),
\end{equation}
 where $\lambda$ and $\mu$ are the two \emph{Lamé} constants.
 This expression matches the one from \textcite{steinke2018}.

\subsubsection{A novel hybrid approach based on a crack boundary condition\label{sec:novel_hybrid_approach}}

Referring to the key requirement defined in \textcite{strobl2015NovelTreatmentCrack} that in a fully damaged state, tensile crack normal stresses and shear stresses along a frictionless crack surface should be zero, \textcite{hu2020PhasefieldModelFracture} developed a stress-based decomposition approach based on a smooth traction-free crack boundary condition.
This approach is similar to the stress-based split by \textcite{steinke2018}, however, instead of considering degradation from the perspective of crack driving stresses in Mode-I, Mode-II and Mode-III, they view degradation as a contact problem.
With $\bm{r}$ as the crack face's normal vector, this results in the following decomposition of the stress tensor:
\begin{align}
    \bm{\sigma}^+ &= \underbrace{\left<\sigma_{rr}\right>_{+}(\bm{r}\otimes\bm{r})}_{\text{tensile normal stress}} + \underbrace{\bm{\sigma} - \sigma_{rr}(\bm{r}\otimes\bm{r})}_{\text{tangential stress}}\\
    \bm{\sigma}^- &= \underbrace{\left<\sigma_{rr}\right>_{-}(\bm{r}\otimes\bm{r})}_\text{compressive normal stress}
\end{align}

The main difference compared to \Cref{eq:total_active_stress,eq:total_passive_stress} is in the treatment of tangential components, which in the case of the crack boundary condition always result in an energetic driving force.
This results in an unrealistic overestimation of the driving strain energy density (\eg $\bm{\sigma}_{tt}$ and $\bm{\sigma}_{ss}$ are considered crack driving).
Therefore, \textcite{hu2020PhasefieldModelFracture} highlight that this decomposition approach should not be used for crack initiation and crack propagation, but should only serve as a boundary condition, which is activated after the phase field variable reaches a certain threshold.

An alternative approach for dealing with the problem of having unphysical crack driving forces is utilizing the properties of the hybrid approach (\Cref{sec:isotropic_anisto}).
As discussed in \Cref{sec:facture_contributing_and_passive_parts}, this method allows arbitrary combinations of energetic driving forces and definitions of the constitutive behavior, given that the solver is based on the staggered approach (see \Cref{sec:Solver}).
Commonly, the isotropic formulation from \Cref{eq:constiso} is used for defining the constitutive behavior, which, however, requires an additional constraint for preventing interpenetration of crack faces.
In order to circumvent this additional constraint, we therefore propose using
\begin{align}
    \bm{\sigma}^+ &= \left<\sigma_{rr}\right>_{+}(\bm{r}\otimes\bm{r}) + \bm{\sigma}_{rt} + \bm{\sigma}_{tr} + \bm{\sigma}_{rs} + \bm{\sigma}_{sr} + \bm{\sigma}_{tt} + \bm{\sigma}_{ss} + \bm{\sigma}_{ts} + \bm{\sigma}_{st} \;\text{and}\label{eq:crack_boundary_active}\\
    \bm{\sigma}^- &= \left<\sigma_{rr}\right>_{-}(\bm{r}\otimes\bm{r})\label{eq:crack_boundary_passive},
\end{align}
which matches the formulation from \textcite{hu2020PhasefieldModelFracture}, for defining the constitutive behavior, where only $\bm{\sigma}^+ $ is degraded.
The crack driving part of the coupled system remains based on the orthotropic stress split derived in \Cref{sec:facture_contributing_and_passive_parts}.
This additionally bypasses the need for considering the crack boundary condition only after the phase field variable reaches a certain threshold, as the driving parts are now rooted in the classic fracture mechanic failure modes.
To conclude, we propose using a stress-based decomposition, where crack driving stresses are identified by fracture mechanics failure modes (\Cref{eq:total_active_stress,eq:total_passive_stress}), in a hybrid-approach, combined with a degradation function, where stresses are degraded, such that the resulting constitutive behavior matches the one of a traction-free crack surface (\Cref{eq:crack_boundary_active,eq:crack_boundary_passive}).
This allows physical, sound estimation of crack driving forces for orthotropic materials and proper modeling of crack faces.

\subsubsection{Application of the stress split in a multi-phase field theory}

\begin{figure}[htb]
    \centering
    \includegraphics{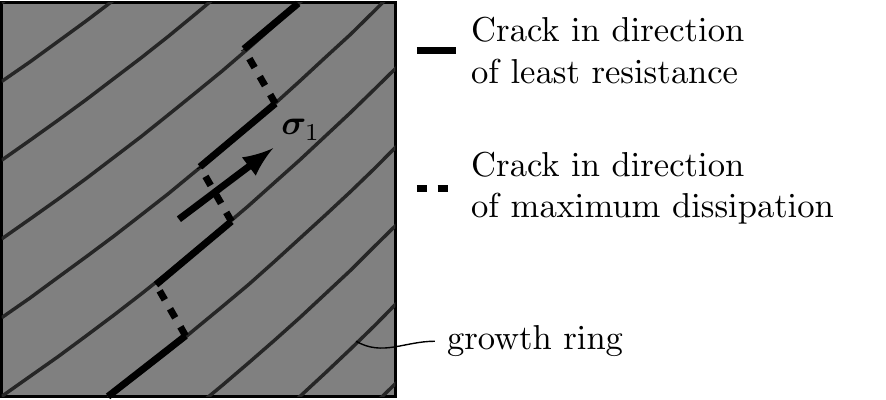}
    \caption{
        Typical crack pattern observed for wood.
        A crack switches from following the direction of least resistance along the fiber, to the direction of maximum total energy reduction perpendicular to the largest principal stress.
        It changes its orientation again, when it hits the next growth ring. \parencite{Smith2003}
    }
    \label{fig:jumping-cracks}
\end{figure}

In highly orthotropic materials, fracture is driven based on two principles \parencite{Smith2003}:
\begin{itemize}
    \item Cracks following the direction of least resistance, defined by the microstructure of the material.
    \item Cracks opening perpendicular to the largest principal stress, thus, leading to a maximum reduction of the total energy.
\end{itemize}
For wood this results in the often observed zig-zag pattern (\Cref{fig:jumping-cracks}), where cracks follow the path of maximum total energy reduction, until reaching a growth ring.
At the growth ring, which is essentially a weak interface, the crack direction changes to the direction of least resistance, which for wood is always parallel to its fiber direction.
So cracks are likely to follow the material's structure, \ie for wood the longitudinal, radial or tangential direction.

The fundamental part of the stress-based decomposition is the definition of the crack coordinate system.
Due to the material characteristics and often observed crack pattern of wood, instead of considering an arbitrary crack face orientation, the identification of three crack coordinate systems defined by $\bm{r}_i$, $\bm{s}_i$ and $\bm{t}_i$ (see \Cref{fig:CSS}) is plausible, where $\bm{r}_1$ is the longitudinal, $\bm{r}_2$ the radial and $\bm{r}_3$ the tangential direction.
This results in three different crack driving energy terms, where according to the principle of maximum dissipation, the failure mode with the highest energy release determines the main cause of failure.

In order to consider the very different fracture toughness and strength of each of the possible types of cracks (longitudinal, radial, tangential), each material axis is related to only a single phase field variable.
This is similar to approaches taken for crystalline materials for considering the influence of cleavage planes \parencite{NGUYEN2017279}.
With this multi-phase field description, only the phase field variable related to driving failure mode is activated for degradation.
In comparison to other multi-phase field models, \eg \textcite{BLEYER2018213, NGUYEN2017279}, this approach results in no coupling of the phase field variables in the constitutive relation, as a single phase field variable is already sufficient to describe the state of a fully developed crack.
Thus, the strain energy terms $\psi^+(\bm{u},\bm{d})$ and $\psi^-(\bm{u},\bm{d})$ in \Cref{eq:phasefieldeq}, are replaced by
\begin{align}
    \psi_i^+(\bm{u},\bm{d}) = \omega_i(d_i) \psi^+_i(\bm{u})\text{ and } \psi_i^-(\bm{u}) = \psi_i(\bm{u}) - \psi^+_i(\bm{u}),
\end{align}
respectively, where $i$ is the index of the defining failure mode.

From a numerical point of view, having no coupling between the phase field \acp{DOF} is favorable.
However, choosing one driving failure mechanism introduces a strong non-linearity, as the strain energy density function depends on $i$, for which the total energy is not a continous function.
This makes solving the problem very difficult, if not impossible.
Therefore, instead of evaluating strong non-linearities (\eg Heaviside functions or the Macaulay brackets) based on the current state variables, they are computed using the deformation and phase field values of the last converged increment.
Given that the increments are sufficiently small, this vastly improves the convergence rate, while leading to similar results.
Our tests showed no significant influence on the obtained solutions, except that the algorithm is more robust.

\subsection{Solver\label{sec:Solver}}

Solving the posed minimization problem is a challenging task, because of the regularized functionals being non-convex in the state variables \parencite{Amor2009}.
Furthermore, to ensure thermodynamical consistency and consider the fully dissipative nature of crack growth, it is necessary to impose irreversibility constraints \parencite{Miehe2010b}.
Solving a non-convex bound-constrained optimization problem for the global optimum is hard; in this case, due to the requirement of a very fine mesh and the resulting problem size, it is almost impossible.
Therefore, searching for local minima is the only option.
Here, mainly two different approaches are pursued:
Application of an alternate minimization scheme (also referred to as the staggered approach) \parencite{Bourdin2000,Miehe2010} and the monolithic approach \parencite{Miehe2010b}.
The consideration of the constraints is not part of these two approaches and is carried out separately.

In the monolithic approach, the problem is solved by using a standard Netwon-Raphson procedure on the full set of state variables.
This can lead to convergence issues, due to the non-convexity of the problem in those variables.
Improvements to these algorithms were proposed, \eg by \textcite{gerasimov2016} in the form of a line search that is performed in case the energy functional increases and more recently by \textcite{kopanicakova2020} through additional application of a thrust region method.
Still, most of the recent works (as well as this) do not rely on the monolithic approach and instead use the staggered one.

\begin{figure}[htb]
    \centering
    \includegraphics{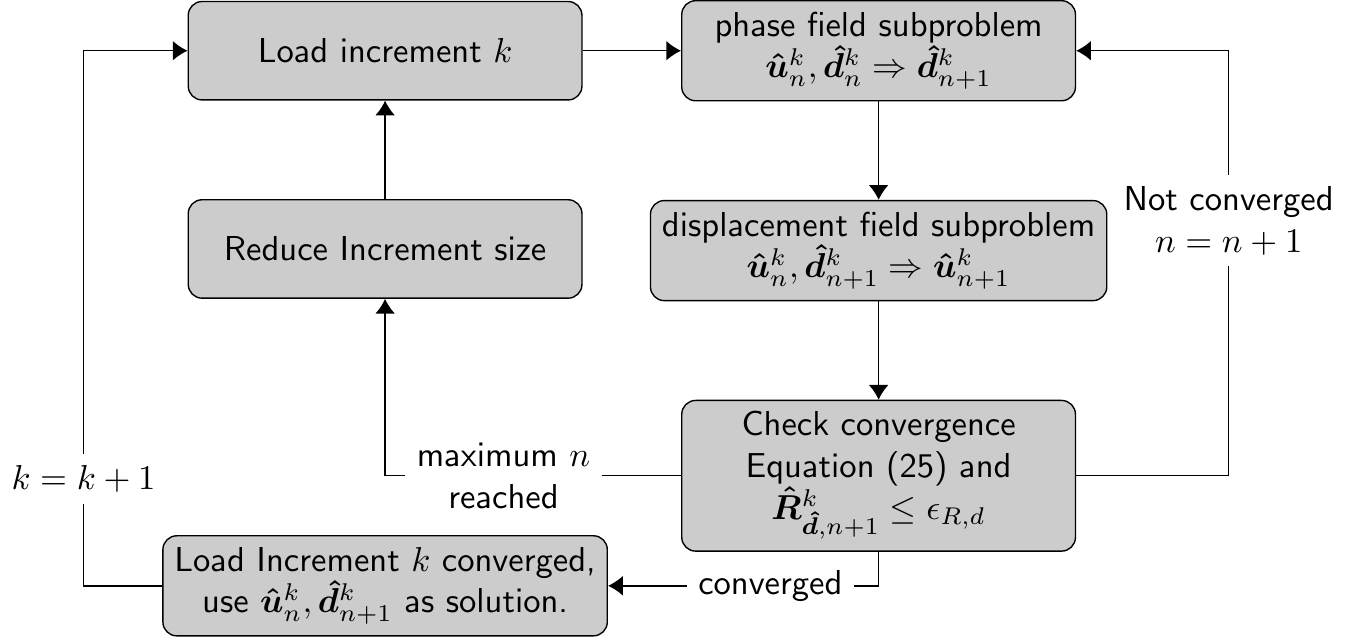}
    \caption{
        Staggered solution process for a single load increment.
        Convergence is assured by imposing a limit on the size of the state variable's increment.
        Additionally, the residual of the phase field subproblem must lie below a threshold.
        The convergence of the overall problem is given by checking whether the last change in the crack phase field only resulted in a converged state in the displacement field.
    }
    \label{fig:staggered-solver}
\end{figure}

The fundamental idea of the staggered approach is to separate the problem into two subproblems, where one is used to determine the displacement field at a constant crack phase field and the other to find the crack phase field at a constant displacement field.
The problems are solved in an alternating manner, such that each is based on the resulting state variable of the other.
One main advantage of the staggered approach is that the subproblems are convex and, thus, the robustness of the solution process is relatively high \parencite{Ambati2015}.
The alternating scheme, however, requires a higher number of iterations compared to the monolithic approach and thus greater computational effort.
Also, convergence criteria are required not only for the overall problem, but for both subproblems.

The algorithmic scheme applied in this work is outlined in the flowchart in \Cref{fig:staggered-solver}.
Similar to what was proposed by \textcite{Amor2009}, the following criteria are used as a convergence measure for the subproblems in iteration $n$ and increment $k$,
\begin{align}
    \sum_{i \in \text{DOFs}}\frac{|\dFE^k_{n+1,i}-\dFE^k_{n,i}|}{|\dFE^k_{n+1,i}|} \leq \epsilon_d \quad
    \sum_{i \in \text{DOFs}}\frac{|\uFE^k_{n+1,i}-\uFE^k_{n,i}|}{|\uFE^k_{n+1,i}|} \leq \epsilon_u \quad
    \text{with}\quad
    \epsilon_u < \epsilon_d.
\end{align}
Here, $\hat{\bullet}$ refers to the discretized representation of the continuous field $\bullet$.
Additionally, the L${^2}$-norm of the crack phase field's residual $\bm{\hat{R}}^k_{\dFE,n+1}$ must be smaller than or equal to $\epsilon_{R,d}$.
Convergence of the overall problem is assured by requiring convergence of the deformation problem and the phase field problem.
If the set of state variables resulting from the last phase field step ($\dFE^k_{n+1}$ and $\uFE^k_{n}$) result in a converged state for the deformation problem, the newly obtained deformation state $\uFE^k_{n+1}$ is ignored (as the phase field subproblem converged for $\uFE^k_{n}$) and $\dFE^k_{n+1}$ and $\uFE^k_{n}$ are accepted as a solution.
The whole algorithm uses an adaptive time increment stepping scheme, such that a better performance can be achieved by having larger increments in less critical regions (\eg linear elastic regime) and smaller increments in critical ones (\eg close to peak load).

As mentioned above, to ensure thermodynamical consistency, an irreversibility constraint on the phase field variable is required.
Additionally, it is also necessary to ensure the bound constraint $d(x) \in [0,1]$.
\textcite{Lorenzis2020} give a quite comprehensive overview of current approaches.
There are various methods and considerations that allow formulating those constraints in various places in the implementation.

Regarding the bound constraint, in \Cref{sec:fundamentals}, the requirement on the degradation function, $\omega'(d_i = 1)=0$, was discussed.
Essentially, this disables the energetic driving force for $d_i = 1$, and therefore, a further development of the related phase field $i$.
Thus, the upper bound is naturally satisfied by the choice of such a degradation function.
Given that there is a solution for enforcing irreversibility of the phase field problem in general, the lower bound is implicitly fulfilled as the crack phase field is initially zero.
Thus, for every subsequent increment the condition $0 \leq \dFE^{0} \leq \dFE^{k} \text{ }\forall\text{ } k$ holds.
Conceptually, two different schemes for the inequality constraint are currently used:
Applying constraints on a local level, like the history field \parencite{Miehe2010} and the penalty method \parencite{Gerasimov2019}, or applying constraints on the global level, by application of Dirichlet-type boundary conditions \parencite{Bourdin2000} or active set methods \parencite{Heister2015, Hintermueller2002}, which are essentially Dirichlet-type boundary conditions on a subset of $\dFE$.
In this work, the computationally less expensive active set reduced space method \parencite{yang2016a}, which is both qualitatively and quantitatively similar to the primal-dual active set method from \textcite{Heister2015}, is used (for a comparison see \Cref{app:activeset}).

A set is said to be \enquote{active} when the inequality constraint is violated.
In \cite{yang2016a} a box constraint solver is discussed, hence identifying two sets, a lower bound active set $\mathcal{A}_\phi$ and an upper bound active set $\mathcal{A}_\psi$.
The problem is, subsequently, solved on the inactive set
\begin{align}
    \mathcal{I}(\dFE^k_{n+1}) = \mathcal{S} \setminus \left(\mathcal{A}_\phi(\dFE^k_{n+1}) \cup \mathcal{A}_\psi(\dFE^k_{n+1})\right),
\end{align}
where $\mathcal{S}$ is the set of all crack phase field \acp{DOF}.
The \ac{DOF} values on the active set are fixed to the boundary values using hard Dirichlet-type boundary conditions. 
Applying this method would allow considering alternative degradation functions for which $\omega'(d_i = 1) \neq 0$, however, in this work no functions of this kind are used.
The active sets are computed as follows\footnote{In their work, \textcite{yang2016a} compare the current \ac{DOF}'s value with the one from the last converged state by equality. Given that they apply a special operator that cuts off values lower than the lower bound and larger than the upper bound, comparing by lower than or larger than, respectively, leads to the same result. We use those operators instead, to make the comparison to the primal-dual active set algorithm from \textcite{Heister2015} more clear.}:
\begin{align}
    \mathcal{A}_\phi(\dFE^k_{n+1}) &= \left\{  i \in \mathcal{S} | \left(\dFE^k_{n+1}\right)_i \leq \left(\dFE^{k-1}\right)_i \text{ and } (\bm{\hat{R}}^k_{\dFE,n+1})_i > 0 \right\}\label{eq:yang-active-set}\\
    \mathcal{A}_\psi(\dFE^k_{n+1}) &= \left\{  i \in \mathcal{S} | \left(\dFE^k_{n+1}\right)_i \geq 1 \text{ and } (\bm{\hat{R}}^k_{\dFE,n+1})_i < 0 \right\},
\end{align}
where the lower bound is the element-wise restriction that the current state variable must be larger than or equal to the state variable from the last converged increment, \ie $k-1$.
These methods also affects the convergence conditions from above, such that the L$^2$-norm of the residual is only computed on the inactive set.
To conclude, we propose using the staggered approach, as it is more robust than the monolithic one, in combination with the active set reduce spaced method, which allows assuring irreversibility without requiring additional terms like penalty functions in the phase field formulation.

The entire code is implemented in Julia \parencite{bezanson2017JuliaFreshApproach}.
For automatically deriving the element stiffness matrices and residual vectors from the energy formulation, the ForwardDiff-Package \parencite{revels2016ForwardModeAutomaticDifferentiation} is used.
Pardiso 6.0 \parencite{deconinck2016NeedlesLargeScaleGenomic, kourounis2018NextGenerationMultiperiod, verbosio2017EnhancingScalabilitySelected} is employed as the sparse linear solver.

\section{Results and discussion\label{sec:results_and_discussion}}

In the following Section, the stress split described in \Cref{sec:facture_contributing_and_passive_parts} is assessed based on two different models: A simple notched plate (\Cref{fig:notchplate-2d}) with varying fiber orientation and a more complex example of a wooden board including a knot (\Cref{fig:simpleknot-3d}).
As the envisioned use of the phase field model described in this work is the application to complex three-dimensional geometries, three-dimensional linear tetrahedral elements are used.
Since the hybrid approach strongly alters the phase field formulation, it can be expected to provide different crack topologies, compared to the variationally consistent formulation (\ie the anisotropic formulation from \Cref{sec:isotropic_anisto}).

For all examples, the material stiffness tensor is defined as $\bm{C}_{\text{LLLL}} = 9000.016,\, \bm{C}_{\text{LLRR}} = 269.384,\, \bm{C}_{\text{LLTT}} = 175.104,\, \bm{C}_{\text{RRRR}} = 480.096,\, \bm{C}_{\text{RRTT}} = 118.528,\, \bm{C}_{\text{TTTT}} = 270.6,\, \bm{C}_{\text{RTRT}} = 32,\, \bm{C}_{\text{LRLR}} = 552,\, \bm{C}_{\text{LTLT}} = 552$, all in $\si{MPa}$.
This resembles the elastic properties of so-called clear wood, describing wood areas without defects and knots.
Following \cite{lukacevic20193DModelKnots}, to the knot in the wooden board example, a stiffness tensor reduced by a factor of $0.5$, compared to the clear wood stiffness tensor, is assigned.
This reduction takes cracks perpendicular to the grain direction, often observed in knots, into account.
The elastic properties are defined in a local cylindrical coordinate system, as is commonly used for describing wood.
$\bm{a}_1$ defines the longitudinal (L) direction, $\bm{a}_2$ the radial (R) direction and $\bm{a}_3$ the tangential (T) direction.

In order to account for the cohesive behavior of wood, the coefficients $a_{1,i}$, $a_{2,i}$ and $a_{3,i}$ in \Cref{eq:Q_a1_a2_a3} are tuned to match a linear softening law.
Based on the analytical solution of a one-dimensional bar problem, \textcite{Wu2017} gives the following definitions:
\begin{align}
    a_{1,i} = \frac{4}{\pi}\frac{l_{ch,i}}{l_i},\; a_{2,i} = -\frac{1}{2}\text{ and } a_{3,i} = 0,\label{eq:a1_a2_a3_softening}
\end{align}
for $\xi = 2$ and $p = 2$, in \Cref{eq:localpart,eq:degrad}, respectively.
$l_{ch,i}$ defines \emph{Irwin}'s characteristic length, given as $l_{ch,i} = \sfrac{E_{0,i} G_{c,i}}{f_{t,i}^2}$, for the $i$-th phase field, and $l_{i}$ is the length scale parameter for the $i$-th phase field, which is chosen to be larger than the effective element size $l_{\text{eff}}$ (third root of the average volume of the finite elements in region of the expected phase field crack).

\subsection{Single edge notched plate\label{sec:res:notchplate}}

\begin{figure}[H]
    \centering
    \includegraphics{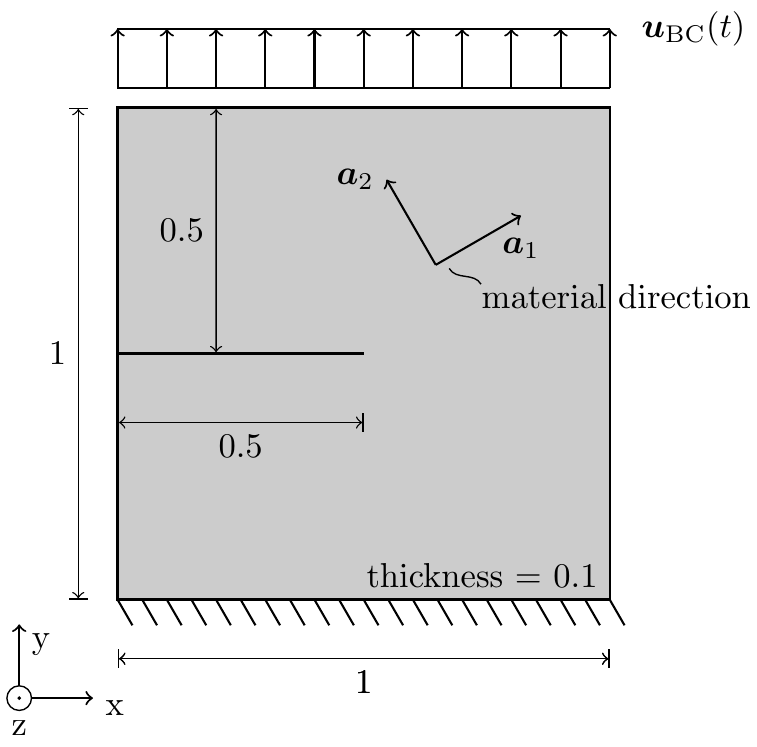}
    \caption{Geometry of the single edge notched plate.
    In plane, the plate is fixed in all directions at the bottom edge and only out of plane across the entire back surface.
       The load is applied in form of a prescribed vertical deformation along the upper edge.
       The fiber direction $\bm{a}_1$ is changed by setting the fiber angle relative to the horizontal direction.
       All measurements are in $\si{mm}$.}
    \label{fig:notchplate-2d}
\end{figure}

The notched plate's geometry is depicted in \Cref{fig:notchplate-2d}.
In plane, it is supported at the bottom edge and out of plane on the entire back surface.
The load is applied in form of a prescribed vertical deformation along the upper edge.
For considering the orthotropic behavior, the fiber direction ($\bm{a}_1$) is changed by setting the fiber angle relative to the horizontal direction, \eg $0^\circ{}$ meaning $\bm{a}_1$ points into the x-direction and $90^\circ{}$ meaning $\bm{a}_1$ points into the y-direction.
The remaining axes are defined such that the tangential direction ($\bm{a}_3$) always points into the z-direction.
The parameters controlling the phase field problem are given in \Cref{tab:param-notchplate}.

\begin{table}[H]
    \centering
    \caption{Defining parameters for the single edge notch plate problem.\label{tab:param-notchplate}}
    \begin{threeparttable}
    \begin{tabular}{!{}lllll!{}}
        \toprule
        $d_i$ & $\beta_i$\tnote{a}  & $G_{c,i}$\tnote{b} & $f_{t,i}$\tnote{c} & $\sfrac{l_i}{l_{\text{eff}}}$ \\
        \midrule
        $d_1$ & 5.0 & 0.05 & 50.0  & 4.0 \\
        $d_2$ & 5.0 & 0.1  & 14.42 & 4.0 \\
        $d_3$ & 5.0 & 0.1  &  7.21 & 4.0 \\
        \bottomrule
    \end{tabular}
    \begin{tablenotes}\footnotesize
     \item[a] Structural tensor scale in \Cref{eq:structural_tensor}
     \item[b] in $\sfrac{\si{Nmm}}{mm^2}$
     \item[c] in $\si{MPa}$
    \end{tablenotes}
    \end{threeparttable}
\end{table}

In order to reduce the computational effort of such problems, often, the mesh density is increased in regions of a priori known crack paths.
As changing the fiber angle is expected to also change the resulting crack topology, the crack paths cannot precisely be known in advance.
Therefore, all models are consistently meshed with the same effective element size over the entire specimen's geometry.
This greatly reduces the influence of the mesh structure on the resulting crack paths.
Initially, seven different element sizes, ranging from a very coarse mesh with 2018 nodes to a very fine one with 144825 nodes, are tested.
The finest mesh results from a characteristic element size of $0.005$, a value which is also used in other publications, \eg by \textcite{hu2020PhasefieldModelFracture}.

\begin{figure}[H]
    \centering
    \includegraphics[width=\textwidth]{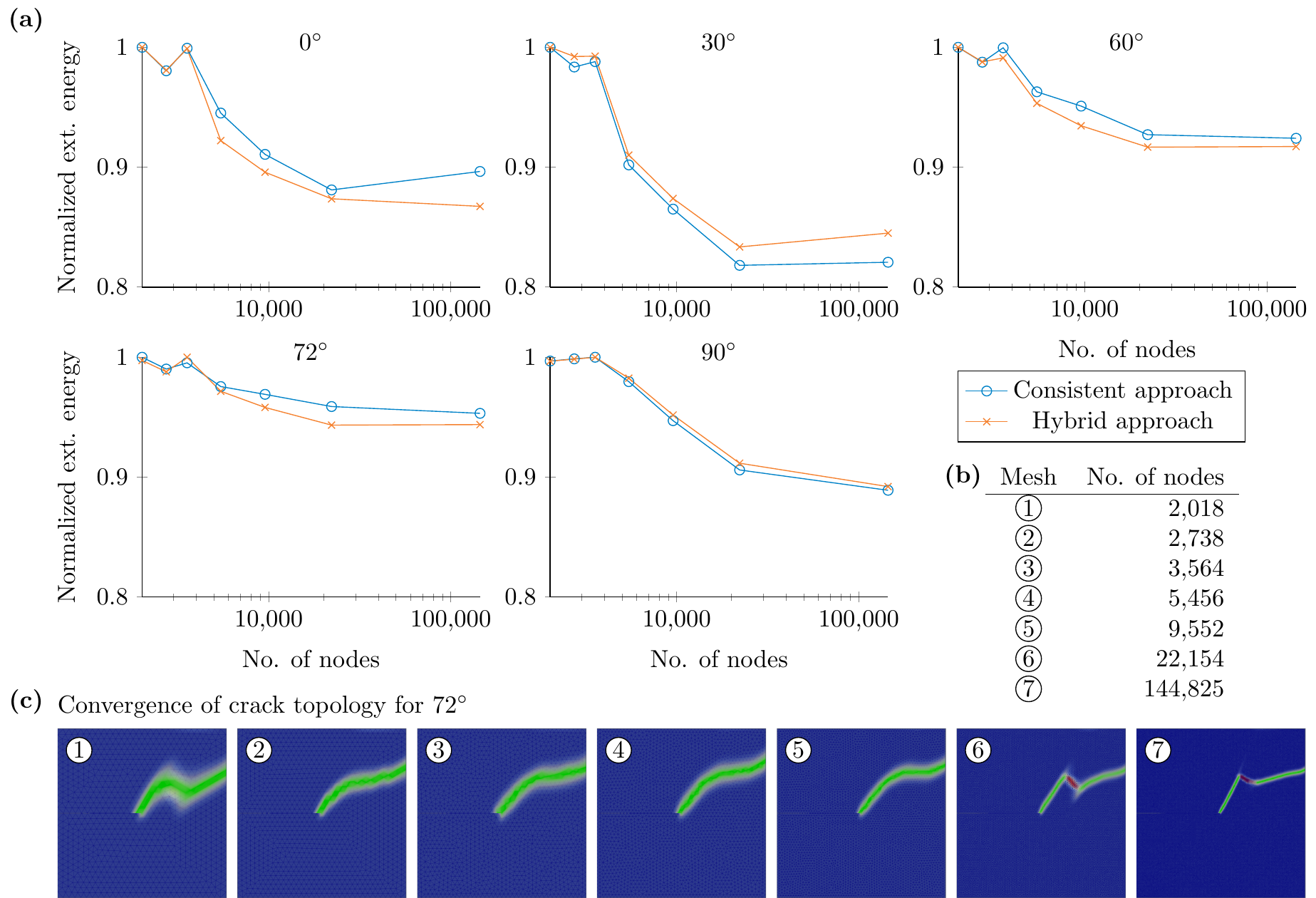}
    \caption{
        Results of the mesh study (total external energy) for seven different meshes with increasing mesh density, for 5 different fiber angles and for both the consistent and the hybrid approach.
        Additionally, the crack topology for a fiber angle of $72^\circ$ for the hybrid approach is depicted for each of the meshes.
    }
    \label{fig:meshstudy}
\end{figure}

The results of this mesh study are shown in \Cref{fig:meshstudy}.
For both the hybrid and the consistent approach, the total external energy, normalized to the maximum value of the external energy for the specific model and the studied fiber angle, is plotted over the number of nodes.
For all five material directions, with increasing number of nodes, the total external energy shows clear convergence against a value that can already be captured well by the two finest meshes (6 and 7).
This is also reflected in the phase field developments in \Cref{fig:meshstudy}~(c), where there is no qualitative difference in the crack topology between mesh 6 and 7, however, a significant change in the failure mode in meshes 1 to 5.
Therefore, only mesh number 7 was used, for the further simulations.
Nevertheless, it should be pointed out that also mesh number 6, though quite coarse compared to common mesh sizes used in literature, could already be used, which would allow a major reduction of the computational effort.
The generally low mesh sensitivity, as also pointed out by \textcite{Yang2019}, is related to the usage of the unified phase field theory from \textcite{Wu2017}, as the regularization parameter is actually considered in the phase field formulation for calibration of the coefficients $a_{1,i}$, $a_{2,i}$ and $a_{3,i}$ in \Cref{eq:a1_a2_a3_softening}, which compensates the size effect resulting from a larger crack phase field.

\begin{figure}[H]
    \centering
    \includegraphics{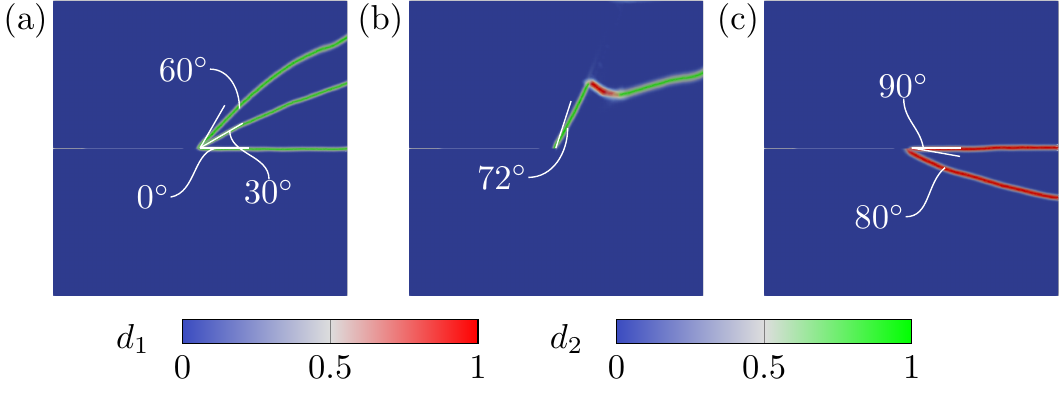}
    \caption{
        Transition of failure modes with increasing fiber incline. Similar failure mechanisms are plotted above each other, \ie \textbf{(a)} shows three different simulations.
        \textbf{(a)} shows only phase field $d_2$, \ie a crack that propagates along the fiber,
        \textbf{(b)} shows the interaction of phase field $d_1$ and $d_2$, \ie cracks that propagates along and perpendicular to the fiber
        \textbf{(c)} shows only phase field $d_1$, \ie a crack that propagates perpendicular to the fiber.
    }
    \label{fig:transition_of_failure_modes}
\end{figure}

As already pointed out, a main question of this work is to evaluate whether the proposed model is capable of considering effects resulting from the material structure appropriately.
With respect to wood, it is of particular interest, whether the commonly observed zig-zag failure pattern (as shown in \Cref{fig:jumping-cracks}), arising from cracks that jump between growth layers, can be modeled.
In order to show this effect, various fiber angles ranging from $0^\circ$ to $90^\circ$ are investigated.
\Cref{fig:transition_of_failure_modes} shows the transition of failure modes, obtained with the hybrid approach, for selected characteristic fiber angles in this range, where the mesh and the load are identical.
In \Cref{fig:transition_of_failure_modes}, qualitatively similar crack topologies are summarized in a graphic to improve comparability.

The results clearly show that at a certain fiber angle, the failure mode switches from a crack driven by stresses perpendicular to the fiber ($d_2$) to a crack driven by stresses in fiber direction ($d_1$).
The first and foremost observation from those results is that by using the proposed hybrid approach, it is actually possible to recover the zig-zag fracture pattern, even if a completely homogenous mesh and material definition is used.
The main influencing factors are the structural tensor, which forces the geometric phase field evolution to stay on planes perpendicular to the crack normal direction, and considering the driving stresses on the fictitious crack face for each likely crack orientation.
This will further be elaborated at the end of this section, where the influence of the hybrid approach on the appearance of this pattern is discussed.

Comparing Figures~\ref{fig:transition_of_failure_modes}~(a) and (c) it becomes obvious that phase field $d_2$ (green) has a stronger tendency to follow the prescribed fiber orientation than phase field $d_1$ (red).
This result suggests that even without explicitly defining a weak interface in-between fibers, cracks perpendicular to the $\bm{a}_2$-direction (\Cref{fig:notchplate-2d}) are influenced by an effect that has a similar impact as a weak interface.
Nevertheless, the cracks do not exactly follow the fiber direction and the stronger the incline is, the stronger the tendency away from this fracture plane.
Theoretically, by increasing the structural tensor scale, one could increase the affinity to this plane, however, closer to the edge of the plate, mixed fracture modes, for example for the $72^\circ$ and the $80^\circ$ model, cause a rather strong deviation.

\begin{figure}[H]
    \centering
    \includegraphics{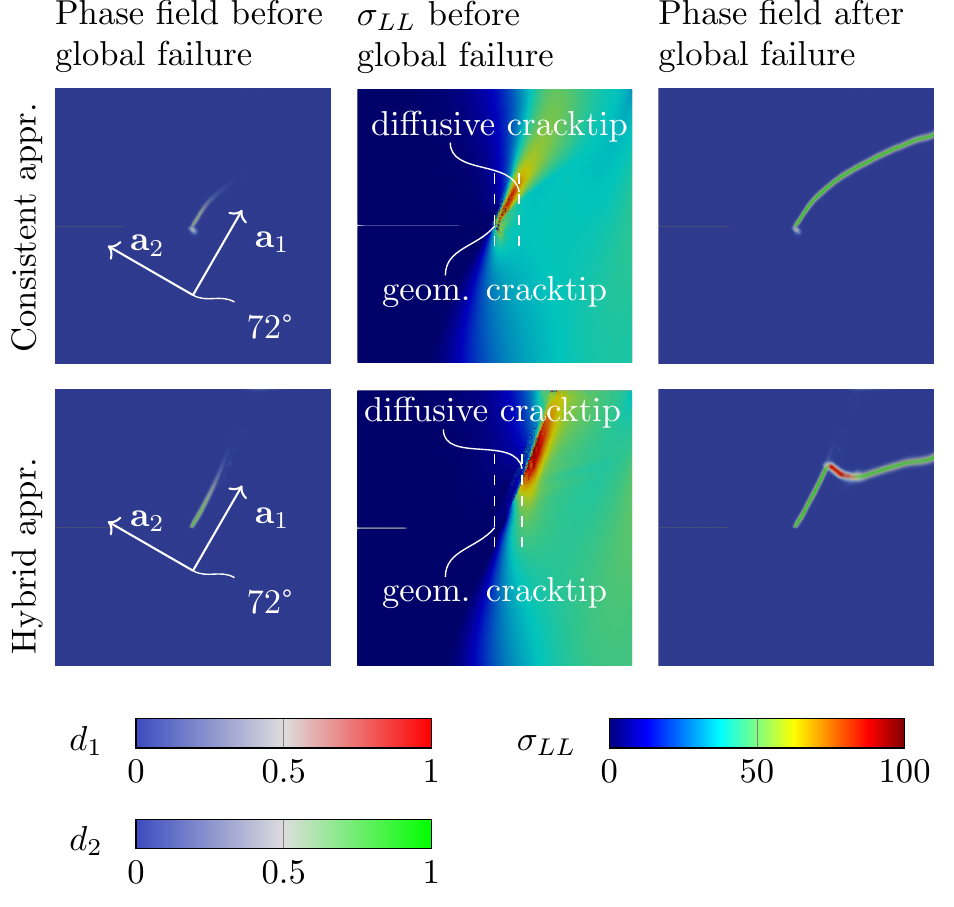}
    \caption{
        Comparison of the hybrid and the consistent approach for two states, one prior to the fully developed phase field and one for the fully developed field.
        The figure shows the main difference in the Mode-I stresses for phase field $d_1$ ($\bm{\sigma}_{LL}$) which leads to the zig-zag pattern for a $72^{\circ}$ fiber incline in case of the hybrid approach.
        Stresses are given in $\si{MPa}$.
     }
    \label{fig:analy_zig_zag}
\end{figure}

For the fiber angles shown in \Cref{fig:transition_of_failure_modes}~(a) and (c), the results of the hybrid approach and the consistent approach agree.
The main difference appeared for the $72^\circ$ case shown in \Cref{fig:transition_of_failure_modes}~(b).
In case of the consistent approach, no interaction occurred, meaning that the zig-zag pattern could not be reproduced (see the final state of the phase field in \Cref{fig:analy_zig_zag}).
The hybrid and the consistent approach differ in how the solid material is degraded, which influences the stress distribution and, thus, the crack driving forces.

To take a close look at these differences, a state prior to the fully developed phase field is compared in \Cref{fig:analy_zig_zag}.
Of primary interest are the stress components which only contribute to the development of one phase field, either $d_1$ or $d_2$, which for this quasi two-dimensional example are the Mode-I stresses, \ie $\bm{\sigma}_{LL}$ for $d_1$ and $\bm{\sigma}_{RR}$ for $d_2$.
As $\bm{\sigma}_{RR}$ is degraded in both approaches, the main difference is in $\bm{\sigma}_{LL}$.
Looking at the stress plot in \Cref{fig:analy_zig_zag}, there is actually a notable difference between the hybrid and the consistent approach.
In the consistent approach the longitudinal stresses are not degraded, thus peaking at the geometrical crack tip from the single edge notched plate.
For the hybrid approach, with the longitudinal stresses being fully degraded, the stress peak moves with the diffuse phase field crack tip, causing a stress state at this location, which favors the evolution of phase field $d_1$.
This is an important finding, because it clearly shows that the usage of the hybrid approach is required in order to recover a zig-zag pattern.

\subsection{Comparison of the hybrid approach and the consistent approach\label{sec:comp_hyb_cons}}

\begin{figure}[H]
    \centering
    \includegraphics{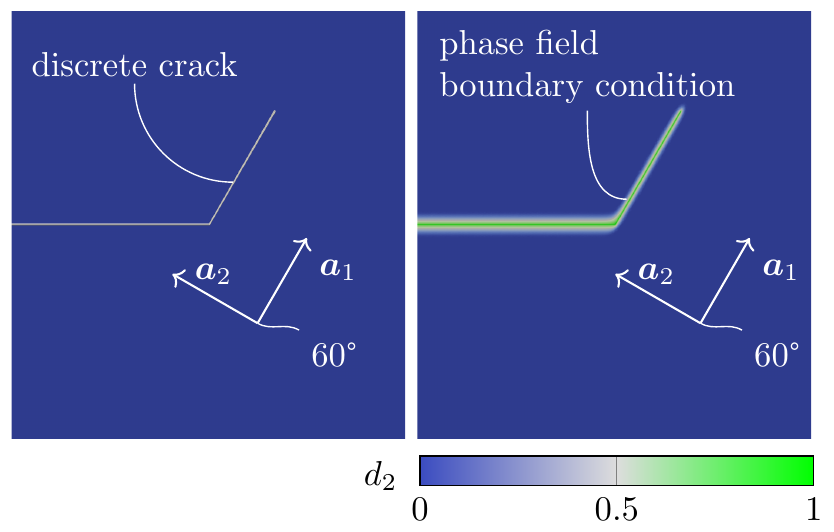}
    \caption{ Two different approaches to compare a phase field crack with a geometrically modeled, discrete crack.
    The phase field crack is computed by prescribing the Dirichlet-type boundary condition $d_2=1$ on the same region as the discrete crack is defined.}
    \label{fig:geom_vs_phase_field}
\end{figure}

While \Cref{fig:analy_zig_zag} gives a strong argument in favor of the hybrid approach, it is still of interest, which of the two approaches is closer to modeling an actual discrete crack.
Therefore, results of both methods are compared with the resulting stress distribution and deformation of a model with a discretely modeled predefined crack.
The two different approaches, including a crack with a kink of $60^\circ$, are shown in \Cref{fig:geom_vs_phase_field}.
To properly compare the two cases, a discrete crack is modeled, and the same crack is modeled by solving the phase field problem for a Dirichlet-type boundary condition, prescribing $d_2 = 1$ on the same region.
Both cracks follow the fiber incline of $60^\circ$.
Subsequently, a linear elastic simulation for the model with the discrete crack and a simulation of the deformation problem considering the phase field distribution from \Cref{fig:geom_vs_phase_field}, using the hybrid and the consistent approach, was conducted.
For all three models the vertical deformation at the top edge is set to $u_\text{y} = 0.1$.

\begin{figure}[H]
    \centering
    \includegraphics{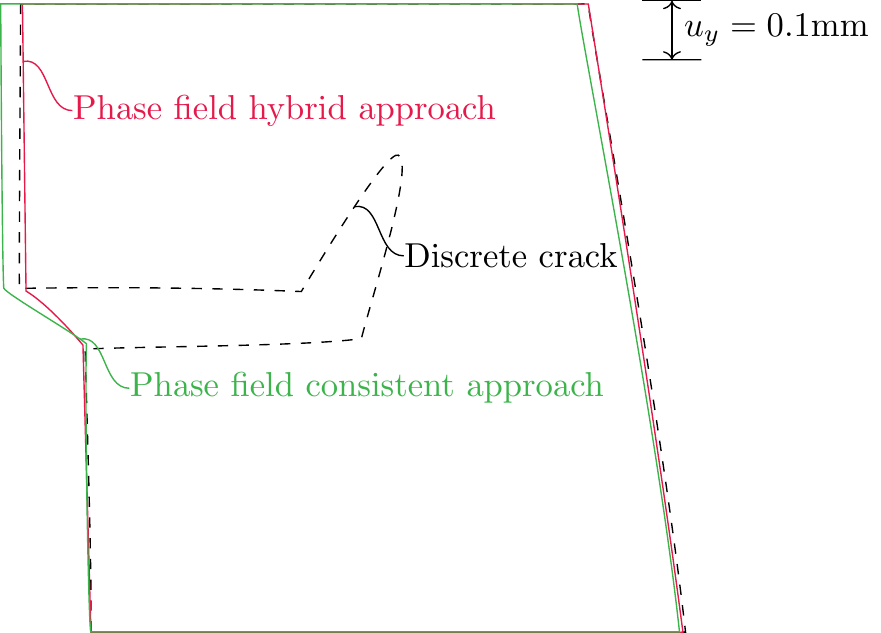}
    \caption{Comparison of the deformation of the outer edges of a model with a discrete crack and two phase field models, using the hybrid approach and the consistent approach.
    }
    \label{fig:deformation_hyb_cons}
\end{figure}

First, the deformation state of the three models is compared.
The results are shown in \Cref{fig:deformation_hyb_cons}, where only the outer edges of each of the notched plates are depicted.
Clearly, as the hybrid approach degrades all elastic components, it resembles the solution of the discrete crack model very well.
In contrast, in the case of the consistent approach, the remaining stresses related to phase field $d_1$ result in a quite large deviation from the discrete crack solution.

\begin{figure}[H]
    \centering
    \includegraphics[width=\textwidth]{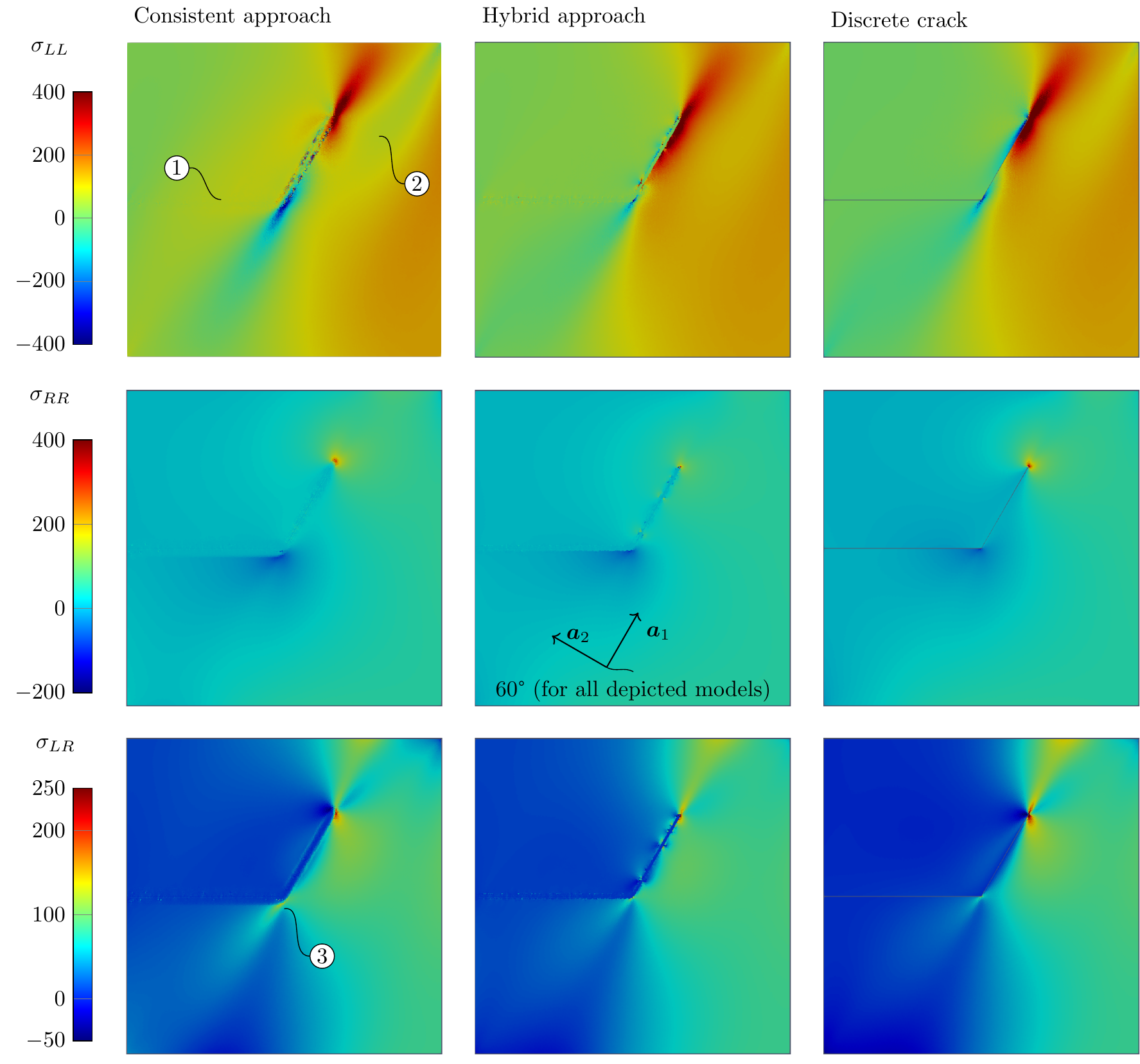}
    \caption{
        Comparison of three stress distributions between the consistent approach and the hybrid approach using a predefined phase field distribution and a discrete crack model.
        \circled{1} shows that longitudinal stresses are still transferred through the crack when using the consistent approach and \circled{2} shows that this does not only affect the region of the crack but also regions further away.
        \circled{3} shows a peak in shear stresses at the kink, when using the consistent approach. Stresses are given in $\si{MPa}$.
    }
    \label{fig:geom_stresses}
\end{figure}

\Cref{fig:geom_stresses} shows a comparison of stress distributions for the -- essentially two-dimensional -- notched plate.
As expected, the main differences between the three models occur for stress components related to the longitudinal direction ($\bm{\sigma}_{LL}$ and $\bm{\sigma}_{LR}$), as those are not fully degraded in the consistent approach.
\Cref{fig:geom_stresses}, region \circled{1} shows that while the specimen is cracked at the given location, there are still stresses transferred through the crack.
Neither the hybrid approach, nor the model with the discrete crack show longitudinal stresses in this region.
Furthermore, this behavior also affects regions away from the crack, \eg in region \circled{2}.
The influence on the remaining stress components is marginal, as $\bm{\sigma}_{RR}$ and $\bm{\sigma}_{LR}$ are fully degraded in both cases.
Nevertheless, in region \circled{3}, the consistent approach shows a peak in shear stresses at the kink which is considerably smaller at the other two models.

Given the similarities of the stress distributions and deformation states, and the hybrid approach's ability to reproduce the zig-zag failure patter found in wood, the results give clear support for using the hybrid approach over the consistent one, when modeling complex material failure.

\subsection{Wooden board with a knot}

\begin{figure}[H]
    \centering
    \includegraphics{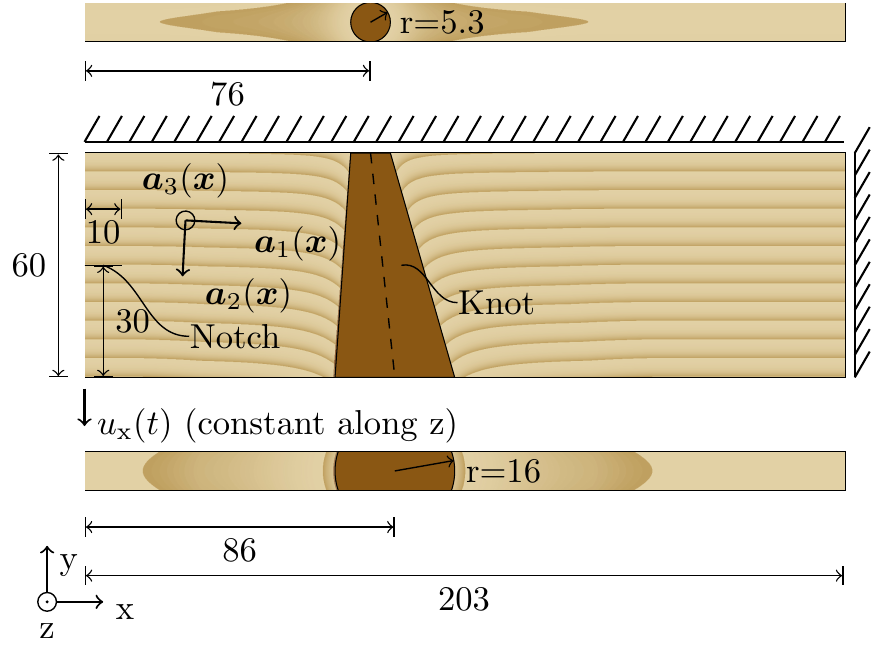}
    \caption{
        Geometry of the wooden board with a single knot.
        The fiber orientation $\bm{a}_1$ (longitudinal), $\bm{a}_2$ (radial), $\bm{a}_3$ (tangential) is prescribed in each integration point and is rendered in the mid-plane of the board.
        A notch is placed in order to control crack initiation.
        The board is loaded at the bottom left edge by prescribing the deformation.
        All measurements are in mm.
    }
    \label{fig:simpleknot-3d}
\end{figure}

\Cref{fig:simpleknot-3d} shows the geometry of the wooden board with a single knot.
The board is supported both at the top and the right face.
The load is applied in form of a prescribed deformation along the bottom left edge.
In order to control crack initiation, the board has a \SI{10}{mm} notch through the specimen.

The fiber orientation $\bm{a}_1$ (longitudinal), $\bm{a}_2$ (radial), $\bm{a}_3$ (tangential), is computed for each individual integration point and defines the local material directions.
In this work, the model from \textcite{lukacevic20193DModelKnots} is used, where a knot is represented by a rotationally symmetric cone and wood fibers are streamlines flowing around an obstacle, which is the knot.
Consequently, the fiber orientation in the LT-plane can be computed using the so-called \emph{Rankine} oval, which describes the fluid flow around an elliptical object.
Additionally, as fibers are situated on so-called growth surfaces, which motivate the cylindrical coordinate system commonly used to describe the elastic properties of wood, the third dimension of the fiber direction vector, the so-called dive angle, can be computed by restricting it to be orthogonal to the LT-plane.
\Cref{fig:simpleknot-3d} also shows a rendering of the fiber course.
The texture is generated based on the previously computed fiber directions on a $0.10\si{mm}$ grid in the mid-plane of the board.

Similar to the notched plate from \Cref{sec:res:notchplate}, the mesh size is homogenous in the left part of the specimen, where the crack will open.
Restricting phase field evolution just to the left part reduces the number of \acp{DOF} and keeping the mesh homogenous reduces the influence of the mesh structure on the crack paths.
The characteristic length of the elements in the phase field activated region is set to $0.75\si{mm}$.
Relative to the volume of this region, this matches the mesh density of notched plate model number 6 in \Cref{fig:meshstudy}, which is in good agreement with the results from an even smaller characteristic length.
The resulting total number of \acp{DOF} is $976,200$.
As outlined in \Cref{sec:comp_hyb_cons}, the hybrid approach is required for properly modeling fracture processes of wood.
Therefore, for simulation of this more complex example, only the hybrid approach is used.
The parameters controlling the phase field problem are given in \Cref{tab:param-simpleknot}.

\begin{table}[H]
    \begin{center}
    \caption{Defining parameters for the wooden board with a single knot.\label{tab:param-simpleknot}}
    \begin{threeparttable}
    \begin{tabular}{!{}lllll!{}}
        \toprule
        $d_i$ & $\beta_i$\tnote{a}  & $G_{c,i}$\tnote{b} & $f_{t,i}$\tnote{c} & $\sfrac{l_i}{l_{\text{eff}}}$ \\
        \midrule
        $d_1$ & 2.0 & 2.0 &  80.0  & 2.0 \\
        $d_2$ & 4.0 & 0.8  & 5.0 & 2.0 \\
        $d_3$ & 2.0 & 0.1  &  3.6 & 2.0 \\
        \bottomrule
    \end{tabular}
    \begin{tablenotes}\footnotesize
     \item[a] Structural tensor scale in \Cref{eq:structural_tensor}
     \item[b] in $\sfrac{\si{Nmm}}{mm^2}$
     \item[c] in $\si{MPa}$
    \end{tablenotes}
\end{threeparttable}
    \end{center}
\end{table}

\begin{figure}[H]
    \centering
    \includegraphics{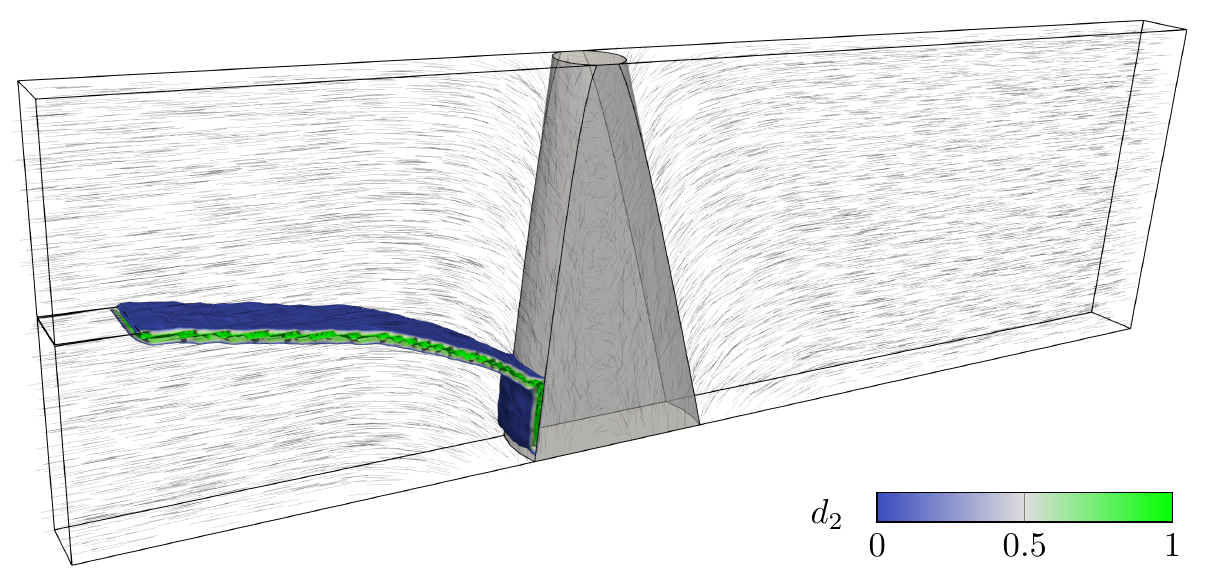}
    \caption{
        Fully cracked wooden board.
        The fiber orientation vector $\bm{a}_1$ is plotted on a uniformly spatially distributed subset of the integration points.
        The crack mostly follows the fiber direction, starting horizontally, tilts and subsequently propagates along the interface region between the clear wood area and the knot.
    }
    \label{fig:simpleknot-cracked}
\end{figure}

\begin{figure}[H]
    \centering
    \includegraphics{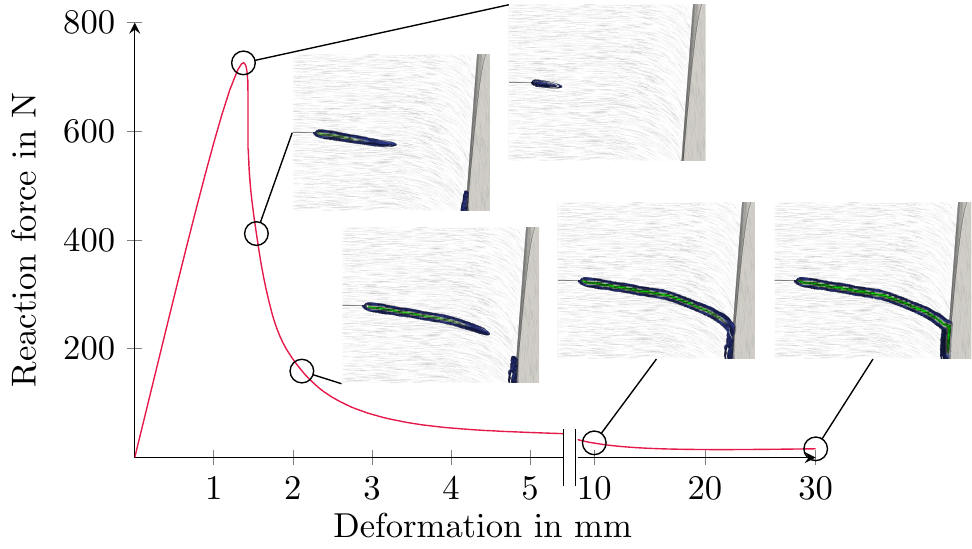}
    \caption{
        Load-deflection plot measured at the lower left edge of the board.
        Additionally, the evolution of the phase field variable $d_2$ is shown for multiple locations along the path.
    }
    \label{fig:simpleknot-xyplot}
\end{figure}

\Cref{fig:simpleknot-cracked} shows the resulting crack path when the specimen is almost fully cracked.
Phase field $d_2$ is visualized using three-dimensional contour lines ranging from $d_2 = 0.0$ to $d_2 = 1.0$ in increments of $0.1$.
The fiber direction is depicted by plotting $\bm{a}_1$ on a uniformly spatially distributed subset of integration points.
Obviously, the varying fiber directions influence the orientation of the crack face.
\Cref{fig:simpleknot-xyplot} shows the evolution of the phase field variable $d_2$.
The crack initially starts with a slight decline and changes its orientation in the vicinity of the knot, where the fibers become parallel to the knot's surface.
When reaching the knot, the crack kinks and follows the weak interface.
It stops propagating close to the lower edge of the board, as the compressive Mode-I stresses in this region do not result in crack driving forces, due to the additive decomposition of the strain energy density term.

\Cref{fig:simpleknot-xyplot} shows the load-deflection plot of the simulation, measured at the lower left edge of the board.
The horizontal axis is split into two differently scaled parts, as the change in the reaction force from \SIrange{0}{3}{mm} is quite large compared to the change from \SIrange{10}{30}{mm}.
Past the initial opening of the crack, the load-deflection plot shows a cohesive behavior during further crack propagation.
The softening effect can be controlled by properly setting $f_{t,i}$ and $G_{c,i}$ in \Cref{eq:a1_a2_a3_softening}.
The reaction force is heavily reduced while the crack propagates along the fiber towards the knot.
With the crack further progressing, tensile and compressive stresses, similar to the bending stresses at the clamped end of a cantilever beam, concentrate at the lower left edge of the knot.
This shift in the stress distribution results in a less stiff response of the system, therefore, larger deformations are required for further crack growth along the weak interface between the clear wood area and the knot.

The simulation of a more complex model considering a realistic fiber course showed that the discussed phase field model is capable of considering the effects resulting from fiber deviations and that sudden changes in the crack face orientation, \eg the kink when the crack reaches the knot, can be modeled.
Furthermore, using the unified phase field theory, adapted to a linear softening law, allows for a cohesive behavior during crack propagation.

\section{Conclusion and Outlook}
The present work addresses the formulation of a phase field model for orthotropic non-brittle materials, able to reproduce multiple, very different failure mechanisms.
In order to extend the phase field method for fracture to support cohesive behavior, the so-called unified phase field theory is applied and tuned to a linear softening law, resembling a cohesive zones model.
Subsequently, a stress-based split for anisotropic materials is derived, which is based on considering Mode-I, Mode-II and Mode-III crack driving stresses on a fictitious crack plane.
The plane's orientations are defined accordingly to material specific fracture planes, in this case for wood: a crack plane perpendicular to the longitudinal, the radial and the tangential direction.
The orientation in which crack growth results in the largest energy dissipation is the driving failure mechanism.

This formulation is coupled in form of a hybrid approach, by separating the energetic driving force term and the actual degradation of the solid.
In this novel hybrid approach, a smooth traction-free crack boundary condition is used, which incorporates a contact constraint and, thus, does not require an additional constraint for preventing interpenetration of crack faces.
This concept is then put into a multi-phase field model, which allows defining a different fracture behavior for each phase field variable individually.
Therefore, very different failure mechanisms can be modeled and described realistically.
In order to consider the effect of the material's structure on the crack paths, a second-order tensor is added to the crack density function, which scales the phase field's gradient on the plane perpendicular to the associated crack orientation vector.
Hence, preferable planes for crack propagation can be defined, \eg a crack perpendicular to the radial direction is likely to propagate along the fiber direction (longitudinal) and less likely to propagate in the radial direction, due to the weak interface between the fibers and the matrix.

The outlined method is tested using two numerical examples of different complexity, both being wooden specimens.
By means of the model of a sinlge edge notched plate, it is shown that changing the fiber orientation leads to different crack topologies, where cracks travel along the fiber when the load direction is in an obtuse angle relative to the fiber direction.
At a certain fiber incline, as expected, the crack kinks and jumps to the next fictitious growth layer, rupturing the fibers in between.
At a sharp angle, the failure mode changes to a crack perpendicular to the fiber orientation.
The simulations showed that such common phenomena of wood (\eg the zig-zag pattern) cannot be recovered when a variationally consistent  approach is used, thus motivating the use of the hybrid approach with a smooth traction free crack boundary condition.
Subsequently, a more complex example of a wooden board with a single knot and a spatially varying fiber orientation was tested.
The model shows that the phase field crack actually follows the curvature of the wood fibers and also allows for sudden changes in the crack face orientation, \eg in the vicinity of the knot where the crack kinks.
Furthermore, the influence of the cohesive behavior during crack propagation can be observed.

This allows the conclusion that the phase field method can be used to model wood failure, as crack phenomena like the zig-zag pattern can be modeled, complex crack topologies can be depicted and cohesive behavior can be considered.
An apparent limitation of this work lies in the formulation of the energetic driving force, which, while allowing the definition of a different fracture characteristic on the level of each phase field variable, allows no distinction between Mode-I, Mode-II and Mode-III.
Thus, always mixed mode failure is assumed.
As this study's focus is on the implementation of a phase field model for wood and investigation of commonly found crack patterns, future research on validating the model with experimental data is needed.
Furthermore, examining more complex examples like wooden boards with multiple knots is of interest.

\section*{Acknowledgement}

This research was funded in whole, or in part, by the Austrian Science Fund (FWF) \verb|Y1093-N30|. For the purpose of open access, the author has applied a CC BY public copyright licence to any Author's Accepted Manuscript version arising from this submission.
The authors also acknowledge gratefully the support by the ForestValue project InnoCrossLam.

\appendix
\section{Comparison of the active set reduced space method and the primal-dual active set method}
\label{app:activeset}

The only difference between the active set reduced space method from \parencite{yang2016a}, and the primal-dual active set method from \textcite{Heister2015}, is in the selection of the active and the inactive set.
Generally speaking, whether a phase field \ac{DOF} is in the active or the inactive set, is determined by two aspects, the current value of the phase field and the current value of the residual.
Adaption to the notation used in this work and rearranging the parts in the primal-dual active set formulation, leads to the following definition of the primal-dual active set:
\begin{equation}
    \mathcal{A}(\dFE^k_{n+1}) = \left\{ i \in \mathcal{S} | c \mathbf{B}_{ii}\left(\dFE^{k-1} - \dFE^k_{n+1} \right)_i  + (\bm{\hat{R}}^k_{\dFE,n+1})_i > 0 \right\},
    \label{eq:heister-active-set}
\end{equation}
where $c$ is a constant larger than 0 and $\mathbf{B}_{ii}$ is the entry of the $i$-th \ac{DOF} in the diagonal mass matrix $\mathbf{B}$.
Given that both $c$ and the mass are strictly larger than 0, their product is as well.
Comparing the two approaches from \Cref{eq:yang-active-set,eq:heister-active-set} leads to the following observations:
\begin{itemize}
    \item If both $\left(\dFE^{k-1} - \dFE^k_{n+1}\right)_i$  and $(\bm{\hat{R}}^k_{\dFE,n+1})_i$ are larger than 0, the $i$-th \ac{DOF} is in both formulations considered to be active.
    \item If both $\left(\dFE^{k-1} - \dFE^k_{n+1}\right)_i$  and $(\bm{\hat{R}}^k_{\dFE,n+1})_i$ are smaller than 0, the $i$-th \ac{DOF} is in both formulations considered to be inactive.
    \item If the sign of the two terms is different, the $i$-th \ac{DOF} is inactive in the method from \textcite{yang2016a}, however, in the method from \textcite{Heister2015} it depends on the choice of the constant $c$, which is not further elaborated in their work.
\end{itemize}

\clearpage

\printbibliography{}
\end{document}